# Investigation of the arrivals' directions differences for consecutive Extensive Air Showers using the data taken by TEL goniometer under GELATICA network

Yuri Verbetsky[1], Manana Svanidze[1], Abesalom Iashvili[1],
Ia Iashvili[2], Levan Kakabadze[1], Nino Jonjoladze[3]

1. E Andronikashvili Institute of Physics under Tbilisi State University, Tbilisi, Georgia
2. The State University of New York at Buffalo, USA
3. J.Gogebashvili Telavi State University, Telavi, Georgia
Email: yuverbetsky@mail.ru; mananasvanidze@yahoo.com

***Abstract***: *The distribution of the angle between the arrival directions of the fronts of consecutive Extensive Air Showers(EAS) with a wide range of a total number of charged particles is studied using the experimental data taken by the EAS 4-detector array "TEL" in Telavi. The station is a part of the* GELATICA *net in Georgia (*GE*orgian* L*arge-area* A*ngle and* TI*me* C*oincidence* A*rray), which is devoted to the study of possible correlations in the arrival times and directions of separate EAS events over large distances. It is shown that the aforementioned directions really are in the essentially independent and can be used for investigation of simultaneous correlations of the EAS arrival times and directions.*

***Keywords***: *Extensive Air Shower, Angular Difference Distribution.*

## *Introduction*

Interest in the investigations of properties of Primary Cosmic Radiation (PCR) has increased significantly over the course of the last decade. The revival of interest is initiated, primarily, by the new possibilities for investigation of the problems of PCR origin and propagation through the space medium. Furthermore, study of PCR observables can provide insight into the Dark Matter (DM) properties.

It is understood that a single PCR particle can interact in the far and near space and generate at one moment several particles or even a jet of particles directed towards the Earth. It can be a normal jet of particles due to normal interaction of the PCR particle with the cosmic medium nucleus. Additionally, the photonuclear disintegration of the PCR nucleus by the solar photons (Gerasimova-Zatsepin effect [1]) can also initiate the effect. Furthermore, the DM decay or annihilation can also lead to such manifestation [2].

After interaction in the atmosphere, such secondary particles with common origin could produce several EAS events that can be observed at a large spatial separation during a short time interval – the manifestation of the so-called "super-preshower" (SPS) [3-5]. Every secondary particle with energy sufficient for the generation of an observable Extensive Air Shower (EAS) is measurable. (Not necessary Very High Energy (VHE) shower). These are rare unexplored composite events and currently only a few examples are known [6-9]. Meanwhile such complex events can provide insight in properties of PCR, such as e.g. the chemical composition, as well as in DM and the interplanetary and interstellar medium (e.g. the medium density distribution). These problems are under worldwide investigation as the most fundamental objectives of Cosmology, Astroparticle Physics and High Energy Physics.

Investigation of PCR in the VHE range is feasible using the terrestrial experiments through the observation of the EAS events. The investigations of this type are being conducted on several huge and expensive installations such as, e.g. the "Pierre Auger Observatory" [10] in Argentina which oc-cupies an area of 3000 km$^2$ and contains 1660 particle detector stations on the Earth surface. However, the detection of rare EAS events, useful for SPS investigation, is possible via much more affordable systems, albeit providing fewer details. Networks of small-scale Cosmic Ray (CR) stations for the detection of EAS events are being developed in America, Europe and Asia (e.g. ALTA, CZELTA, HiSPARC, LAAS, Showers of Knowledge, etc.) The GELATICA is one of such networks in Georgia.

A station of this type consists of a few straddled detectors that register the passage of the charge particles in the EAS front. Through coincidence of signals from detectors within a sufficiently small time interval, the arrival time of an EAS is obtained. The stations (nodes) of the network are usually placed at local universities or high schools, with the internodal distances in the range from kilometers to thousands of kilometers. The UTC (Coordinated Universal Time) instants of EAS arrival times are determined through the Global Positioning System (GPS) at every station. These timestamps are used to obtain the time correlation of the EAS events detected by different stations and thereby to investigate possible signal of the super-preshowers [6-9]. Different networks of this type are handling the worldwide exchange of data obtained separately; however a uniform method is under discussion.

The above-discussed configuration of globally spaced network of the small scale stations allows investigation of the spatiotemporal correlations of individual EAS events in the domain of PCR-DM under the condition of great distances between the points of showers detection.

It should be noted that these correlations cannot be observed within any of the existing large observatories as their baselines do not extend to more than ~50 km. Moreover, these observatories are not designed for detection of the EAS of low and intermediate energies, which are most likely in the SPS problem. That is why a new Cosmic-Ray Extremely Distributed Observatory (CREDO [3]) has been organized recently. The forming of CREDO global collaboration for observation of global-scale cosmic ray ensembles is under progress and our network GELATICA is a member of this collaboration.

The stations of the GELATICA network possess the capability of EAS arrival direction estimation beyond the standard capability of the arrival time measurement. The posterior joint comparison of the direction and arrival time values measured at the remote locations would allow more reliable selection of the SPS events, rejecting EAS pairs with the accidentally close arrival times.

Investigation of a joint distribution of time intervals and direction differences between the consecutive EAS events at remote locations is the final objective. The first step towards this aim is to study joint and marginal distributions of both these quantities at a single location.

The distribution of time intervals between the consecutive EAS events arrivals at the remote locations have been previously studied by our team [9]. In this article we study the distribution of difference between the consecutive EAS pairs' arrival directions at a single location in the terms of the quantity defined in the following.

***The choice of estimate for the EAS pairs' arrivals directions differences.***

We need to define an estimate for the proximity between the measured arrival directions of two EAS fronts. When the directions of the fronts are fully determined in 3D space, the cosine of the angle $\gamma$ between these directions is used: $\hat{u} \equiv \cos(\hat{\gamma}) = \hat{\mathbf{n}}_1 \cdot \hat{\mathbf{n}}_2$, $|\hat{\mathbf{n}}_1| = |\hat{\mathbf{n}}_2| = 1$, where $\hat{\mathbf{n}}_1$ and $\hat{\mathbf{n}}_2$ are directional unit 3-vectors of the EAS fronts.

EAS arrival direction is typically measured by the so-called "flat goniometers" [11]. This is a device used in our setup and its arrangement allows estimation of only two "horizontal" components $n_x, n_y$ of the EAS front directional axis together with associated uncertainties. (It is assumed that the goniometer is installed on a horizontal level). The uncertainty in the measurements can cause violation of unit modulus for the estimate of the directional vector. This can often lead to imaginary solution for the third "upright" component of the direction $n_z = \sqrt{1-\left(n_x{}^2 + n_y{}^2\right)}$ .

Given the limitations in the measurements, there are at least two options to define estimate sensitive to difference in the directions of EAS fronts.

**1**. The squared difference of two 2-vectors $\vec{n}_1$ and $\vec{n}_2$ using the only two measured projections of the directional 3-vectors onto the horizontal plane:

$$s=(\vec{n}_1-\vec{n}_2)^2=(n_{1x}-n_{2x})^2+(n_{1y}-n_{2y})^2; \qquad |\vec{n}|\neq 1 \tag{1.1}$$

This variable $s$ uses only directly measured quantities by a flat goniometer [11] with no additional assumptions made. However, the limitation here is that the method can not be used if the measurements of $\vec{n}_1$ and $\vec{n}_2$ are obtained by two goniometers located in points with different geographical coordinates (longitudes and latitudes). In this case the two local horizontal planes are not parallel and the estimate (1.1)loses meaning.

The comparison of directions measured in the remote locations must be performed in a single coordinate system common for every goniometer located at any point on the Earth surface. So every direction measured at any location must be represented in the terms of this single system. But such coordinate transformation needs use of all three vector components. That is why the next case seems to be a reasonable compromise:

**2.** The cosine of the *conditional* angle between two directions

$$u=\mathbf{n}_1\cdot\mathbf{n}_2 \tag{1.2}$$

Here the *explicitly reconstructed* directional 3-vectors are used:

$$\mathbf{n}=\begin{pmatrix} n_x \\ n_y \\ \sqrt{1-n_x{}^2-n_y{}^2}\end{pmatrix}; \qquad |\mathbf{n}|=1 \tag{1.3}$$

This definition too applies the condition of unit length of directing 3-vectors. Thus it does not take into consideration the fluctuation of the lengths of complete 3-vectors, but it preserves the information on the fluctuations of the horizontal components. Meanwhile the definition (1.2) (1.3) provides *the measure between the spatial directions*. Thus it is suitable for estimation of difference between the directions of EAS arrivals obtained by the flat goniometers at remote geographic locations.

In the rest of the paper we study distribution density of angular quantity $u$ as defined by (1.2) under the assumption that two EAS-s and their directions are independent.

***Determination of distribution of cosines of angles between the reconstructed arrival directions of independent Extensive Air Showers***

Let us assume that EAS flow does not depend on the shower's arrival azimuth. This assumption allows the easy-to-use application of polar coordinate system for description of the measured horizontal 2-vector of the shower's arrival direction

$$\vec{n}=\begin{pmatrix} n_x \\ n_y \end{pmatrix}=\beta\begin{pmatrix}\cos(\varphi)\\ \sin(\varphi)\end{pmatrix}, \qquad \beta=\sqrt{n_x{}^2+n_y{}^2}, \qquad \beta\neq 1 \tag{2.1}$$

Under the assumption of the azimuthal independence of the showers' flow it is acceptable to represent the complete distribution of the directional 2-vectors $\vec{n}$ as a product of two independent distributions: that of the azimuth angle $\varphi$ and the so-called [12-14] "zenith separation" $\beta$ (2.1):

$$f_{\beta\varphi}(\beta,\varphi) \propto \Theta(\varphi)\Theta(2\pi-\varphi)\times f_\beta(\beta)\cdot\Theta(\beta)\Theta(1-\beta) \tag{2.2}$$

Here $\Theta(x)$ is the unit step function. In the rest of the paper the "$\propto$" sign indicates the equality accurate within the normalization factor.

The estimate of the difference in the directions (1.2) is a function of two azimuth values and two zenith separations:

$$u=\mathbf{n}_1\cdot\mathbf{n}_2=\mathbf{n}(\beta_1,\varphi_1)\cdot\mathbf{n}(\beta_2,\varphi_2)=U(\beta_1,\varphi_1,\beta_2,\varphi_2) \tag{2.3}$$

Here the reconstructed directional 3-vectors (1.3) are expressed in terms of the polar components of the measured horizontal 2-vectors:

$$\mathbf{n} = \begin{pmatrix} \beta\cdot\cos(\varphi) \\ \beta\cdot\sin(\varphi) \\ \sqrt{1-\beta^2} \end{pmatrix}, \quad 0 \le \beta \le 1,\ \ 0 \le \varphi < 2\pi$$

Thus the function (2.3) can be specified as $U(\beta_1,\varphi_1,\beta_2,\varphi_2) = \sqrt{1-\beta_1^2}\sqrt{1-\beta_2^2} + \beta_1\beta_2\cdot\cos(\varphi_1-\varphi_2)$.

Under the assumption (2.2) for the distributions of EAS directions and accepting the hypotheses of mutual independence of the two directions let us define the distribution of $u$ values as given in (2.3) by means of integrations by two pairs of two EAS directional parameters $\varphi,\ \beta$.

$$f_u(u) \propto \iint d\beta_1\, d\varphi_1 \iint d\beta_2\, d\varphi_2\, f_{\beta\varphi}(\beta_1,\varphi_1)\cdot f_{\beta\varphi}(\beta_2,\varphi_2)\,\delta\big(u-U(\beta_1,\varphi_1,\beta_2,\varphi_2)\big)d\varphi_1$$

The mutual independence of distributions of azimuths and zenith separations implies detachment in both integrations' pairs. Then both integrations by azimuths allow complete evaluation:

$$K(u,\beta_1,\beta_2) \propto \int_0^{2\pi}\int_0^{2\pi} d\varphi_1 d\varphi_2\, \delta\big(u-U(\beta_1,\varphi_1,\beta_2,\varphi_2)\big)$$

Introducing new variables $\xi = \varphi_1-\varphi_2$ and $\eta = \varphi_1+\varphi_2$, the δ-function can be written as:

$$\delta\big(u-U(\beta_1,\beta_2,\xi)\big) = \frac{\delta\left(\xi - \arccos\left(\dfrac{u-\sqrt{1-\beta_1^2}\cdot\sqrt{1-\beta_2^2}}{\beta_1\cdot\beta_2}\right)\right)}{|\beta_1\beta_2\cdot\sin(\xi)|}$$

Consequently the kernel $K(u,\beta_1,\beta_2)$ is evaluated as:

$$K(u,\beta_1,\beta_2) \propto \left(\frac{1}{\beta_1\beta_2}\right)\cdot\Theta\left(1-\left|\frac{u-\sqrt{1-\beta_1^2}\cdot\sqrt{1-\beta_2^2}}{\beta_1\cdot\beta_2}\right|\right)\cdot\frac{2\pi-\omega(u,\beta_1,\beta_2)}{\left|\sin\big(\omega(u,\beta_1,\beta_2)\big)\right|};$$

$$\text{here}\quad \omega(u,\beta_1,\beta_2) = \arccos\left(\frac{u-\sqrt{1-\beta_1^2}\cdot\sqrt{1-\beta_2^2}}{\beta_1\cdot\beta_2}\right).$$

Thus the distribution of $u$ value, which measures the difference between two *reconstructed* directions (1.3), gets the structure

$$f_u(u) \propto \int_0^1\int_0^1 K(u,\beta_1,\beta_2)\ f_\beta(\beta_1)\cdot f_\beta(\beta_2)\,d\beta_1\,d\beta_2$$

It depends on the shape of the distribution $f_\beta(\beta)$ of measured zenith separations only, as the kernel $K(u,\beta_1,\beta_2)$ is fully defined by the assumption of azimuth isotropy of the EAS flow.

***Observations***

For the final calculation of the distribution $f_u(u)$ of cosines of *conditional* angle between two *reconstructed* directions, for the purpose of this study, we make use of the known distribution $f_\beta(\beta)$ of measured zenith separations obtained from the same TEL goniometer data [13] in year 2014. The shape of this distribution is shown in figure 1.

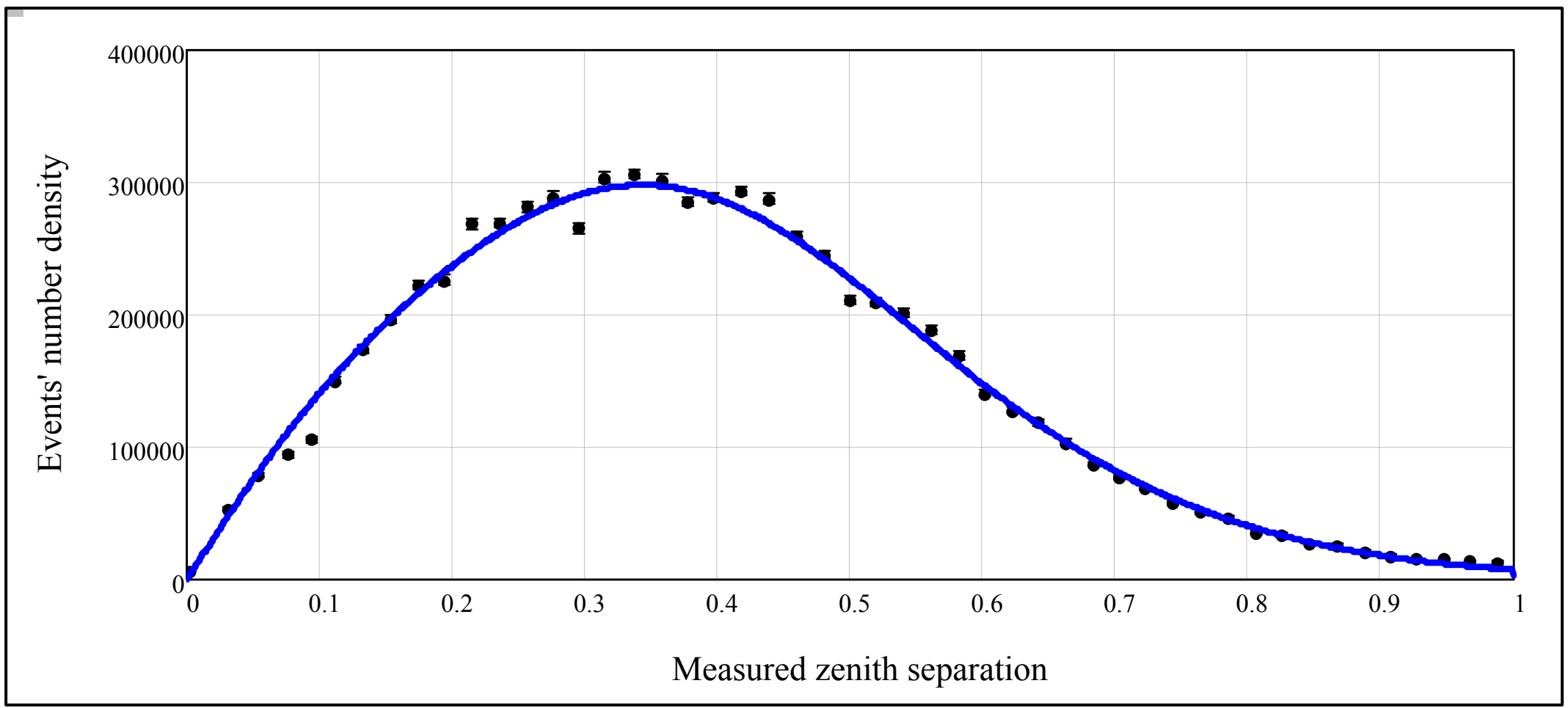

**Figure 1** Distribution density $f_\beta(\beta)$ of the measured zenith separations $\beta$ using data [13] taken by the TEL goniometer in 2014.

The data collected on $41\,870$ pairs of consecutive showers obtained by the TEL goniometer during year 2019 are used next. The horizontal components $n_x, n_y$ of 2-vector of the shower's arrival direction are estimated for every EAS, and the corresponding reconstructed 3-vectors (1.3) are calculated. The value $u$ (1.2) is evaluated for every consecutive pair of showers.

Figure 2 shows distribution density $f_u(u)$ normalized to the total number of EAS pairs used, with the density histogram of $u$-values obtained by the above-described data.

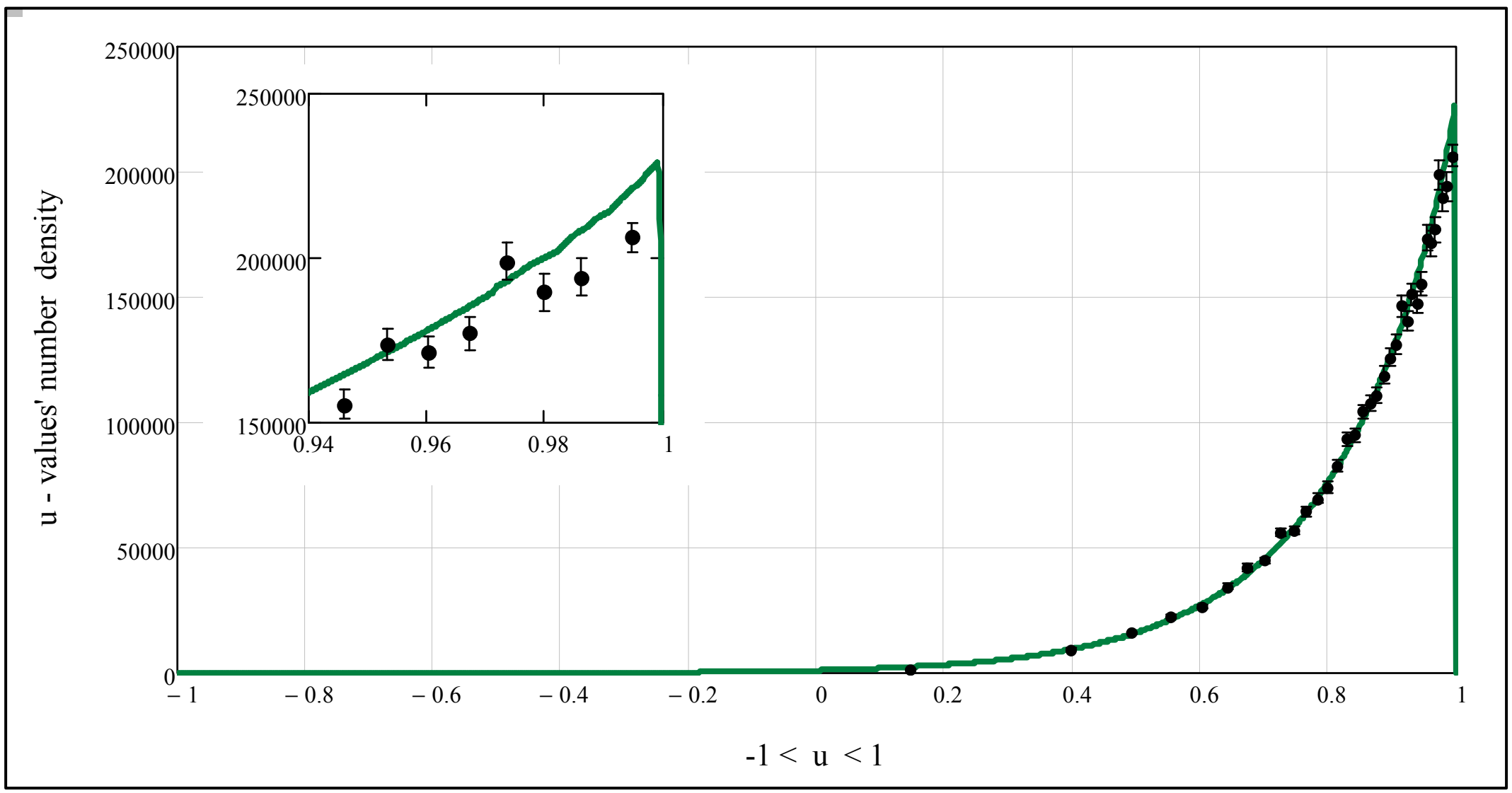

**Figure 2** Distribution density $f_u(u)$ of cosines of conditional angle between the consecutive EAS pairs' arrival reconstructed directions.
Predictions and observed data are shown.

It can be seen that the distribution of cosines of angles between the EAS arrivals' reconstructed directions agree well with the assumption of mutual independence of successive EAS events. Obtained $\chi^2/NoF$ is somewhat larger then 2, but is likely due to statistical fluctuation.

The main objective of the investigations by GELATICA network is identification of the super-preshowers events. Therefore the consecutive EAS pairs with the smallest difference between their arrival directions are of the special interest. Here, the conditional angle $\gamma = \arccos(u)$ of the showers' arrival directions difference provides good sensitivity.

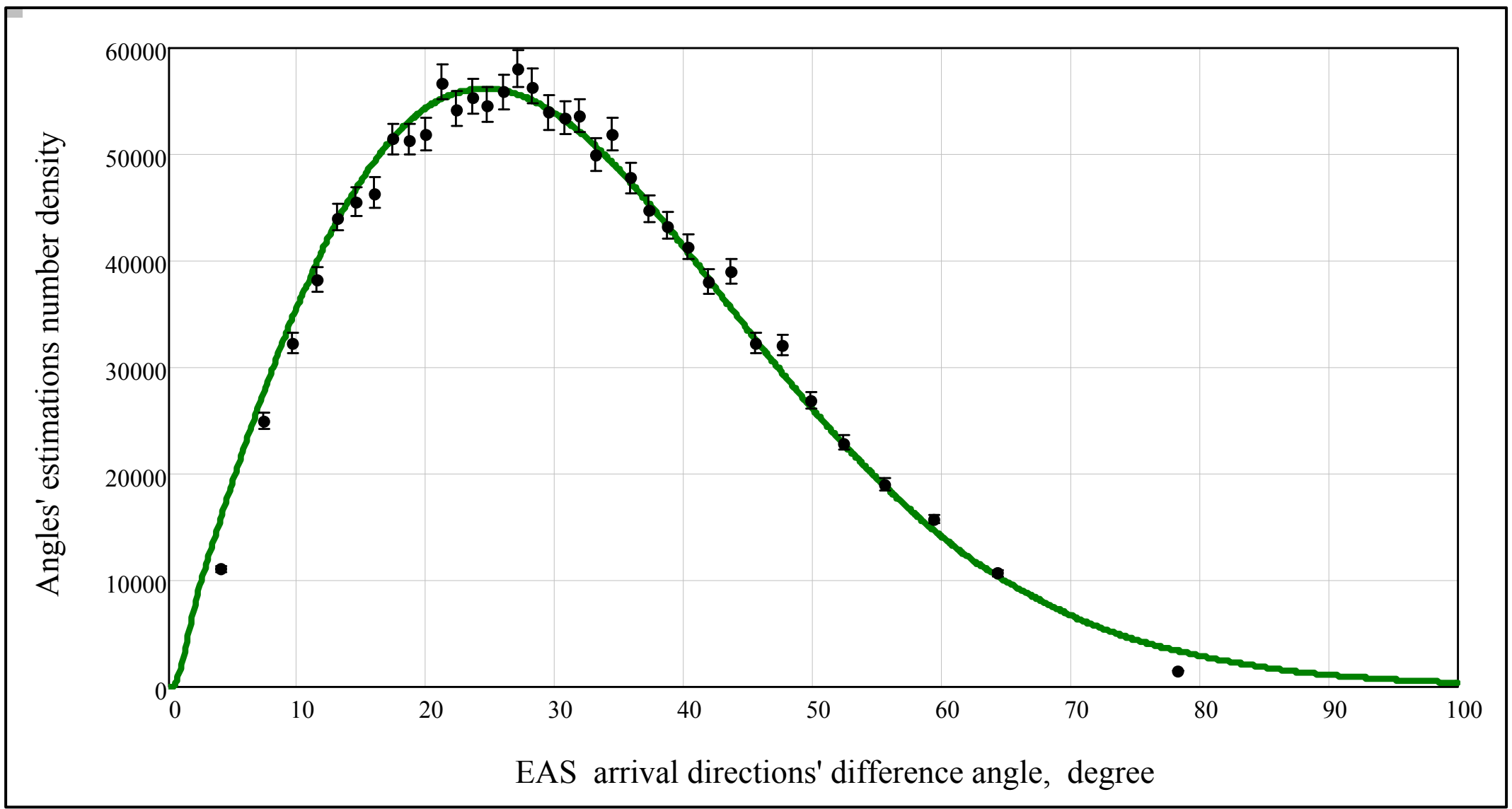

**Figure 3** Distribution density $f_\gamma(\gamma)$ of conditional angle between the consecutive EAS pairs' reconstructed arrival directions. Predictions and observed data are shown.

The distribution density of the $\gamma$ angles $f_\gamma(\gamma) = \sin(\gamma) \cdot f_u(\cos(\gamma))$ normalized to the total number of consecutive EAS pairs used is shown in figure 3 with the density histogram of the estimated $\gamma$-angles obtained by the above-mentioned data. Only a small fraction of the total EAS pairs populate small values of difference angles. Thus the possible super-preshowers events are expected to have rather little background in the conditional angle estimations.

### *Conclusions*

We present a study to show that the arrival directions of the consecutive showers are largely independent. The results are based on data taken in 2019 by TEL goniometer under GELATICA network. Small observed deviation of the data from the expectation under the assumption of the independence is likely of statistical nature.

We conclude that joint use of the time intervals between the consecutive showers together with the conditional angles between their arrivals' directions, calculated from the reconstructed 3-vectors, shell provide more reliable identification of the special EAS pairs observed at remote EAS goniometers, which allow their interpretation as being generated by the required super-preshowers.

### *Acknowledgements*

The authors are grateful to other current and former members of our group for their technical support. We are especially thankful to our colleagues working now abroad. This work was supported by the Georgian National Science Foundation subsidy for a grant of scientific researches #GNSF/ST06/4-075 (№ 356/07) and by Shota Rustaveli National Science Foundation, Project #DI/6/6-300/12.

***References***

**1** Gerasimova N.M., Zatsepin G.T.: "Disintegration of cosmic ray nuclei by solar photons"; Sov. Phys. JETP 11, 899

**2** Warsaw Workshop on Non-Standard Dark Matter: http://indico.fuw.edu.pl/conferenceDisplay.py?confId=45

**3** Cosmic-Ray Extremely Distributed Observatory (CREDO): http://credo.science/

**4** O. Sushchov, P. Homola, N. Dhital, et al: "Cosmic-Ray Extremely Distributed Observatory: a global cosmic ray detection framework"; arXiv:1709.05230 [astro-ph.IM]

**5** P. Homola, G. Bhatta, Ł. Bratek, et al: "Search for Extensive Photon Cascades with the Cosmic-Ray Extremely Distributed Observatory"; arXiv:1804.05614 [astro-ph.IM]

**6** Atsushi Iyono, Hiroki Matsumoto, Kazuhide Okei et al: "Parallel and Simultaneous EAS events due to Gerasimova-Zatsepin effects observed by LAAS experiments"; proceedings of the 31st ICRC, Łódź; http://icrc2009.uni.lodz.pl/proc/pdf/icrc0941.pdf

**7** Karel Smolek, Filip Blaschke, Jakub Čermak et al: "ALTA/CZELTA – a sparse very large air shower array: overview of the experiment and first results"; proceedings of the 31st ICRC, Łódź; http://icrc2009.uni.lodz.pl/proc/pdf/icrc1300.pdf

**8** F. Blaschke, J. Čermák, J. Hubík et al: "CZELTA: An overview of the CZECH large-area time coincidence array"; Astrophys. Space Sci. Trans., 7, pp.69–73; www.astrophys-space-sci-trans.net/7/69/2011

**9** Yu Verbetsky, M Svanidze, A Iashvili, E Tskhadadze and D Kokorashvili: "First results on the spatiotemporal correlations of the remote Extensive Air Shower pairs". 23rd ECRS (and 32nd RCRC) Moscow J. Phys.: Conf. Ser. 409 (2013) 012085; http://iopscience.iop.org/article/10.1088/1742-6596/409/1/012085/meta;jsessionid=89322C00C25D03FA0B0D63C56F0B2333.c5.iopscience.cld.iop.org

**10** Pierre Auger Observatory: http://www.auger.org/

**11** M.S. Svanidze, Yu.G. Verbetsky: "Improved Method of the Extensive Air Shower Arrival Direction Estimation"; http://arxiv.org/abs/0804.1751v1

**12** Yuri Verbetsky, Manana Svanidze, Abesalom Iashvili, Levan Kakabadze. "Extensive Air Showers' Arrival Direction Distribution by TBS Array", International Journal of High Energy Physics. Vol. 1, No. 4, pp.49-54, http://www.sciencepublishinggroup.com/journal/archive.aspx?journalid=124&issueid=-1

**13** Manana Svanidze, Yuri Verbetsky, Ia Iashvili, et al: "Angular distribution of extensive air showers by TEL array under GELATICA experiment"; GESJ: Physics 2016 №1(15) pp.54-62; http://gesj.internet-academy.org.ge/download.php?id=2740.pdf

**14** Yu. G. Verbetsky, M. S. Svanidze, A. Iashvili, I. Iashvili, L. Kakabadze "Extensive air showers' arrival direction distribution by the TSU array under GELATICA experiment"; GESJ: Physics 2018 №2(20) pp.43–55; http://gesj.internet-academy.org.ge/download.php?id=3175.pdf